\documentstyle[epsfig]{article}

        \newcommand{\bea}{\begin{eqnarray}}
        \newcommand{\eea}{\end{eqnarray}}

\title{ %
         \rightline{DESY 95-069}
         \rightline{hep-ph/9504297}
         \bigskip
         { \bf                       On particle--like jets}
      }
\author{                             Hermann He{\ss}ling
       \\
                        Deutsches Elektronen--Synchrotron (DESY) \\
                        Notkestra{\ss}e 85, D--22603 Hamburg, Germany
       }

\begin{document}

\maketitle

\begin{abstract}
Under which conditions does a jet appear as a particle--like signal
from the hidden realm of quarks and gluons?
Motivated by this question jet clustering conditions
are formulated,  in order to characterize jet clustering algorithms,
which can be used  for a
determination of   particle--like jets.
Jets are understood as particle--like, if they  behave like free particles.
The simplest solution to the jet clustering  conditions leads to
a new jet algorithm: a Lorentz invariant generalization
of the  JADE algorithm.
It is found that  this generalization
amplifies hadronization effects in certain phase space regions
in such a way, that
hadronization models  might
become testable  in jet physics  at
the electron--proton collider HERA.
Moreover, a method is suggested, which can be used
at HERA, in order to determine
a region in the phase space, where
hadronization effects from the proton remnant are small and
where parton jets are particle--like.
\end{abstract}

\section{Introduction}

Quarks and gluons  do not exist:
they cannot be investigated experimentally as free particles.
While the mass of a free particle is an observable,
the mass of a quark is not  well defineable.
%
%
Only gauge invariant collections of quarks
and gluons are physically meaningful.
Because of this  confinement property it is difficult to
obtain observable predictions from QCD.
Various strategies were developed to obtain  approximate predictions.
A perturbation theory at short distances can be formulated,
since QCD is asymptotically free.
It  provides deep insights in   the dynamical structures of QCD.
But the hadronization of quarks and gluons cannot be explained
within the perturbative approach. For a description of hadronization
phenomenological  models have to be used.

In high--energy collision experiments hadron jets, i.e. spatially isolated
clusters of hadrons,  are observed.
An interesting question is: under which conditions does
the
direction of hadron jets agree with
the direction of partons from perturbative QCD?
An attempt to answer this question, has to take
into account that partons are colored and hadrons are colorless.

Monte Carlo event generators were developed, which
connect perturbative QCD and hadronization models,
like LEPTO\footnote{
  LEPTO links
  leading order QCD matrix elements (ME) of the  order
  $\alpha_s$, parton shower (PS) and the LUND string fragmentation model.}
\cite{lepto}.
They can be used to study the influence of  hadronization.
Parton jets and   hadron jets
are determined event by event
by applying  a jet algorithm to the outgoing partons of perturbative
QCD and the hadrons from  the phenomenological hadronization
model, respectively. The differences between parton jets and  hadron jets
are caused by hadronization and are called hadronization effects.
Hadronization effects depend strongly
on the jet algorithm, see  e.g. \cite{gut}, \cite{kramermagn},
\cite{op1},
\cite{bethke},
\cite{catani}, \cite{ingelman}.

A jet algorithm is a tool, to analyze the hadronic final state and to
provide a basis for testing the underlying theory.
There are many jet  algorithms.
One reason to create a jet algorithm is,
to make the calculation of special aspects of the
theory possible, see e.g.
the Geneva algorithm \cite{bethke} and the kT--algorithm \cite{catani}.
 Jet algorithms
 motivated by QCD--calculations
 allow a detailled view into certain regions of
 the phase space, but they might not be
 suited equally in different regions.
 For example, in the deep inelastic scattering (DIS) region
 at the electron--proton collider
 HERA it is expected, that the kT--algorithm
 describes jets the better, the more
 they are separated from
 the proton remnant, i.e. the more
 the proton remnant ``factorizes''
 from other jets.
In section~2 a new jet algorithm is derived,
not from  explicit QCD--calculations, but from
general physical considerations.
This jet algorithm
is a Lorentz invariant generalization of the JADE
algorithm. It can be applied
to all areas of jet physics, as long as the jet masses are small.
Using the DJANGO Monte Carlo\footnote{
   DJANGO is an interface between HERACLES, a Monte Carlo
   for  deep inelastic lepton--proton scattering including
   electroweak radiation corrections \cite{heracles}, and LEPTO.
 }
\cite{spiesberger}
hadronization effects
are compared in section~3
for the JADE algorithm
and the new jet algorithm.
A variable is introduced, which assigns a Lorentz invariant
distance between  the jets and the incoming proton.
It is shown that jet masses influence strongly the hadronization effects,
if the jets are close
(with respect to this variable)
to the incoming proton.
The new jet algorithm seems to be a sensitive tool to test
hadronization models in jet physics.


\section{Formulation of jet clustering conditions}

We are interested in
a description of jets, which behave like free particles
and which we call {\em particle--like  jets}.
In this section we formulate four conditions for  jet
clustering algorithms.
They lead  to a new jet algorithm, which can be used to
determine
particle--like
jets.

Clustering algorithms have been developed
in \cite{l},
\cite{d}, \cite{dmb}. We concentrate on clustering algorithms
of the following kind  \cite{jade}.
Consider  a collision event consisting  of  $N$ hadrons with
momenta
$p_1, \dots, p_N$ and  assign a {\em distance}
$
   d_{ij}, 1 \le  i < j \le N,
$
to all pairs of hadrons.\footnote{
  The jet clustering algorithm ARCLUS
  assigns a distance $d_{ijk}$ to all
  triples of hadrons \cite{arclus}.
  }
 By a renumbering of the hadrons one
always manages that $d_{1N}$ is the smallest of all
distances. If $d_{1N}$ is smaller than some reference distance
\bea
          d_{1N}^2 &\le& y_{\rm cut}M^2,
\label{distc}
\eea
recombine  hadron $N$ with  hadron $1$ and
assign a 4--tuple
to the recombined hadrons. We  call
the 4--tuple {\em pseudo--momentum} and
characterize it by the symbol
$
        p_1 \oplus p_N.
$
(In general it is not justified to consider $p_1\oplus p_N$ as a
momentum.)
$M$ is  called {\em reference mass}
and $y_{\rm cut}$ is called {\em jet resolution parameter}.
This clustering procedure is repeated until none of the
distances $d_{ij}$ fulfils the inequality (\ref{distc}).
The remaining hadron clusters define the jet content of
the event. These jets are called {\em hadron jets}.

In perturbative QCD
jet algorithms can
be applied to partons
in order to  obtain results, which are free of
infrared and collinear singularities
(see e.g. \cite{kramer}).
 The resulting jets are called {\em parton
jets}.

Commonly used  clustering algorithms (see e.g.
\cite{gut}, \cite{kramermagn},
\cite{op1},
\cite{bethke},
\cite{catani}, \cite{ingelman}
and references therein) differ by
the reference mass $M$, the
distance $d_{ij}$, and the
recombination prescription $p_i \oplus p_j $.

\vspace*{6mm}

Hadrons, produced in a
high--energy particle collision, propagate almost independently.
Consequently, the forces between hadron jets are small.
Parton jets, on the other hand,
are influenced by hadronization effects because of the
color forces between the partons.
If the color forces between  parton jets are small,
parton jets behave like free particles
and
are characterizable by kinematical
quantities.

We assume that a momentum is sufficient to describe a
particle--like jet and that internal degrees of freedom, like spin,
can be neglected.
An event consisting of $n$ jets, where each jet is particle--like, is called
particle--like $n$--jet.
Energy--momentum conservation suggests, to identify
the total momentum of a particle--like $n$--jet with
the total momentum of its hadrons or partons, respectively.
What can be said about the  momentum of an individual  jet
within a particle--like $n$--jet event?
Consider an event consisting of 2 hadrons with the momenta
$p_1$ and $p_2$.
If the 2 hadrons are recombined,
the only way to guarantee momentum conservation is, to assign the
momentum $p_1+p_2$ to the 1--jet.
By induction one is led to the

\begin{itemize}
 \item[A)] {\em Recombination condition:
The pseudo--momentum has to be identified with the sum of the
momenta to be recombined}
\bea
        p_i \oplus p_j &=& p_i + p_j .
\label{merg}
\eea

\end{itemize}

In the current calculations of perturbative QCD parton jets are
massless
(see e.g. \cite{kramer}, \cite{graudenz} and references therein).
Recombination prescriptions were proposed in the literature
which
guarantee that the masses of the hadron jets are also  zero,
e.g.
$
    p_i \oplus p_j = ( |\vec p_i + \vec p_j |,\vec p_i + \vec p_j )
\label{bloe}.
$
These prescriptions are arbitrary, because
there are uncountable many ways to map
the hadron momenta into  massless hadron jets.
This arbitrariness it  reduced, but not completely, if one
restricts to
 recombination prescriptions which lead to small
hadronization effects.  One is still left with
the problem that the recombination prescription is not universal,
since it
depends on a test variable, which is used as a measure
for the strength of the hadronization effects.
Moreover,
if
the pseudo--momentum is not simply the
sum of the two recombined momenta,
$p_i \oplus p_j \neq p_i+p_j$,
the physical meaning
of the pseudo--momentum is not clear and
in particular one cannot expect, that the direction of a hadron jet
can be identified with the direction of a parton jet.

Another way to treat the masses of jets, is to compare
data with a more complex evaluation of QCD matrix elements.
This, of course, requires the calculation of higher than leading-order
          corrections.

\vspace*{6mm}

The reference distance on the right hand side of  (\ref{distc})
is defined as the product
 of the reference mass squared $M^2$ and the jet resolution parameter
$y_{\rm cut}$.
If the jet resolution parameter $y_{\rm cut}$ is chosen sufficiently large,
the right hand side of (\ref{distc}) can be made  larger than
every  distance squared $d_{ij}^2$ and, consequently,   all hadrons
momenta $p_1, \dots, p_N$ are finally recombined into a 1--jet
with
the momentum $p_1 + \dots + p_N$.
   In the hadronic center of mass frame a 1--jet is at rest
   and its energy is the mass of the hadronic final state.
Let $y_{\rm cut,triv}$ be the smallest $y_{\rm cut}$ value, at which
the hadrons of an event are recombined into a 1--jet.
Because a  1--jet is a trivial jet,
it does not make sense to consider values of the
 jet resolution parameter higher
than $y_{\rm cut, triv}$, i.e.
the jet resolution parameter has to be restricted to the
interval
$
        0 \le y_{\rm cut} \le y_{\rm cut, triv}.
$
By a suitable rescaling of the
reference mass $M$, one always achieves that
$y_{\rm cut,triv}=1$.

\begin{itemize}
\item[B)] {\em Boundary condition:
The reference mass $M$ has to be chosen such that,
 for $y_{\rm
cut}=1$,
all hadrons are recombined into a 1--jet
and that, for $y_{\rm cut}<1$, only higher order $n$--jets, $n>1$,
are generated. }
\end{itemize}

A simple solution to this condition is obtained by an
identification
of the square of the reference mass with the
sum of the squares of all distances between the hadrons
\bea
          M^2 &=& \sum_{i<j} d_{ij}^2.
\label{mass}
\eea
This choice is not unique. For example one could weight each
distance
$d_{ij}$ in (\ref{mass}) with a factor $w_{ij} \ge 1$,
but $w_{ij}=1$ is distinguished, since it
leads to the smallest reference mass $M$.

The reference mass $M$ changes, in general, from clustering step
to clustering step.
On the one hand this might cause problems in theoretical
calculations, but on the other hand there is a lot of freedom
to tune the reference mass, e.g. by means of the weights $w_{ij}$.

\vspace*{6mm}

Finally we formulate two
conditions on the distance $d_{ij}$.

Almost all commonly used jet algorithms
are not Lorentz invariant.
Consequently the jet predictions of these algorithms
depend  on the chosen
reference frame.
This causes experimental problems, if the laboratory frame
does not agree with the reference frame, suited for a non--Lorentz invariant
jet algorithm.
Due to the strong energy imbalance between the
electrons and protons at HERA, the laboratory frame does not
belong to one of the theoretically preferred reference frames,
like the hadronic center of mass frame or the Breit frame.
Besides the experimental errors in the measurement of
jets, a considerable source of  additional uncertainty is introduced at HERA
through
an event by event measurement
of certain kinematical
observables. These kinematical observables  are needed in order to
determine the transformation between the
laboratory frame and the reference frame of a non--Lorentz invariant
jet algorithm.
The additional uncertainty might reduce the applicability of a
non--Lorentz invariant jet algorithm.

Lorentz invariant jet algorithms are
very convenient, at
least from an experimental point of view, since
they can be used in the laboratory frame.
The laboratory frame is
the only natural frame to study detector acceptance effects.
To find the proper acceptance cuts is one of the most important and
most difficult experimental tasks.
But also from a theoretical point of view Lorentz invariant
jet algorithms are preferred, if one is interested in particle--like
jets and if one assumes, that particle--like jets
belong to the observables of QCD.
Since QCD  is Lorentz invariant\footnote{
     A short discussion of the problem, whether the Lorentz invariance
     might be broken in the vacuum of gauge theories, can be found
     \cite{haag}.},
the determination
of particle--like jets  cannot depend on the
inertial reference frame where the jet algorithm is applied.
This leads us to the

\begin{itemize}
\item[C)] {\em Lorentz invariance condition:
The distance $d_{ij}$
has to be independent of the inertial reference frame.}
\end{itemize}

\vspace*{3mm}

In QCD the radiation of collinear partons
is more probable
than the radiation of  non--collinear partons, if one lets aside
infrared partons.
%
%
Collinear partons are  recombined with the highest priority,
if one assigns a vanishing distance between them.
This motivates the

\begin{itemize}
\item[D)] {\em Collinearity condition:
The distance between collinear jets should vanish}
\bea
       d_{ij}  &=& 0, \;\;\;\; {\rm if}\;
        p_j^\mu = \lambda p_i^\mu.
\label{colli}
\eea
\end{itemize}

Collinearity is understood as
collinearity of 4--momenta.
 The
collinearity factor $\lambda$ has to be positive, because
of the positivity of the energy.
%
%

\vspace*{6mm}

Let us analyze the consequences of the clustering conditions C) and D).

Because of the Lorentz invariance condition C)
the distance $d_{ij}$ is
a function of the invariants $p_i^2, p_j^2, p_ip_j$.
We are therefore able to expand
\bea
     \frac{   d_{ij}^2}{M^2} = a_o + \sum_{k=1}^3 a_k r_k
                   + \frac{1}{2!} \sum_{k,l=1}^3 a_{kl} r_kr_l + \dots
\label{tay}
\eea
into powers of the invariants
\bea
         r_1 = \sqrt{\frac{p_i^2}{M^2}},\;
         r_2 = \sqrt{\frac{p_j^2}{M^2}},\;
         r_3 = \sqrt{\frac{p_ip_j}{M^2}}  .
\label{ri}
\eea
For a moment let us assume that the jet  masses
$m_j, m_i$
are small compared to the reference mass $M$.
If in addition the hadrons are almost infrared
($p_i \sim 0$),  or  almost collinear
($p_j^\mu \sim \lambda p_i^\mu$
 with a  collinearity factor $\lambda$ of
the order of 1),
one also has $p_ip_j \ll M^2$.
Then it is allowed
to stop
 the expansion (\ref{tay}) at the quadratic level.
It turns out that the
clustering condition D) determines
the distance $d_{ij}$ up to an overall normalization factor $c=a_{33}$
\bea
        d_{ij}^2 = c \left( (p_i+p_j)^2 - (m_i+m_j)^2 \right),
\label{dist}
\eea
i.e.
we are able to derive
 a distance from the
clustering conditions C) and D), which is unique (up to normalization)
for small jet masses.

The normalization factor $c$ has to be positive, otherwise the
distance $d_{ij}$ is not a real number (to see this, use $p_ip_j \geq m_im_j$).
If the reference mass is identified with (\ref{mass}), the
normalization
factor $c$ does not influence the inequality (\ref{distc}) of
the clustering algorithm. In this case one can set $c=1$.

In the limit of massless jets
the distance (\ref{dist}) becomes proportional
to the distance of the  JADE algorithm \cite{kramerII}, \cite{kramer},
 \cite{jade}
\bea
       d_{ij}^2 \rightarrow c 2  E_i E_j (1-\cos \theta_{ij}), \;\;
    {\rm as}\; m_i,m_j \rightarrow 0.
\nonumber
\eea
The distance $d_{ij}$ would not become the JADE distance
in the limit of small jet masses,
if the invariants $r_i$ in (\ref{ri}) were defined without the
square root.

During a clustering procedure the mass of a
particle--like jet increases.
Practically  one  arrives at
jet masses which are large
compared to the masses of the individual hadrons
within a jet and not negligible against the reference mass.
Beyond
   the first clustering step
   the JADE algorithm violates Lorentz invariance more and more,
   because its distance
   $     2  E_i E_j (1-\cos \theta_{ij})$
   is Lorentz invariant only for massless jets.

The distance of the E--clustering algorithm \cite{kramerII}, \cite{kramer},
$d^2_{ij}=(p_i+p_j)^2$,
is Lorentz invariant, but it does not
fulfil the collinearity condition D), if the jets are massive.

 It should be noticed that
(\ref{dist}) does not define a distance in the classical sense.
If it were a classical distance the triangle inequality
$ d_{ij}+d_{jk}\geq d_{ik}$ would hold, but the triangle inequality  is
violated because of the collinearity conditon D).
To see this, consider e.g. three hadrons with momenta $p_1,p_2,p_3$
and assume that two of them are collinear: $p_3 = \lambda p_2$;
then the sum of two of the distances,
$d_{13} + d_{23} = \sqrt{\lambda}d_{12}$, is  smaller than  the
third distance, $d_{12}$, for all $\lambda < 1$.

Combining (\ref{mass}) and (\ref{dist})
the square of the reference mass becomes (with c=1)
formally similar to the distance (\ref{dist})
\bea
      M^2 = W^2  -(m_1 + m_2 + \dots )^2
 \label{smallmass}
\eea
where
$W^2=(p_1 + \dots + p_N)^2$ is
the square of the mass of the hadronic final state.
We see that the
reference mass $M$ can be chosen smaller than the
mass of the hadronic final state $W$ without contradicting
the boundary condition B).

If the jet masses $m_i,m_j$
are large,
higher order corrections
 in  (\ref{tay}) have to be taken into
account and the distance $d_{ij}$ is no longer unique because
the condition C) and D) are
not violated if
the right hand side of (\ref{dist}) is multiplied by
any positive function
$f(m_i,m_j, p_ip_j)$.

The distance (\ref{dist}) with $c=1$,
 the reference mass (\ref{smallmass}) and
the recombination prescription (\ref{merg}) define
 a new jet clustering algorithm,
which is the simplest algorithm allowed in the setting.
This jet algorithm can be applied to all areas of jet physics,
as long as the jet masses are small compared to the
reference mass. It is  suited for the investigation
of hadronization effects around  remnants
and the determination of particle--like parton jets,
 as is demonstrated in the
next section in the case of deep inelastic electron--proton scattering.
Theoretical calculations based on
this jet algorithm
are well defined, since
collinear and infrared
singularities   are
identifiable. For certain theoretical problems, e.g.
the exponentiation problem,
 it might be
indicated, to modify the distance (\ref{dist})
by higher order mass corrections and/or
 to use a different reference mass $M$.

\section{Hadronization effects}

In this section we present a  study of hadronization effects for the
JADE algorithm, called  jet algorithm 1,
and its simplest generalization to massive jets,
as presented in the last section, called  jet algorithm 2.
The jet algorithm 1 is not Lorentz invariant and needs the
specification
of an inertial system. We consider it in the hadronic center of mass frame.
Using the DJANGO Monte Carlo \cite{spiesberger} we generate
5000 deep inelastic scattering events of 26.7 GeV electrons and
820 GeV protons.
DJANGO is able to take into account corrections from the bremsstrahlung
of the electron.
Let $p_{\rm prot},p_e, p_{e'}, p_\gamma$ be the momenta of the incoming
proton, the incoming electron,
the scattered electron and the bremsstrahlungs photon,
respectively, and
let us define $Q^2=-q^2$ and $Q^2_{\rm had}=-q^2_{\rm had}$, where
$q = p_e - p_{e'}$ and $ q_{\rm had}= q-p_\gamma$.
The events are generated with
$Q^2 > 80 \; GeV^2$. This choice
restricts
$
  x_{\rm Bj} = Q^2_{\rm had}/(2 p_{\rm prot} q_{\rm had})
$
to $x_{\rm Bj}> 10^{-3}$.

LEPTO offers different options to study various aspects of the
electron--proton scattering:
a matrix element (ME) option to generate
1+1 or 2+1 parton jets
 \footnote{ The proton remnant is counted as ``+1'' jet.}
according to leading order QCD matrix elements of order $\alpha_s$,
a parton shower (PS) option,
                                             and a fragmentation
option for the generation of  hadrons out of partons based on the
LUND string
fragmentation model \cite{jetset}. Since these options can be combined
freely, partons at two different levels can be extracted from LEPTO:
 a ME--parton level
and a PS-parton level. In the following we identify the
parton level with the ME--parton level.
In LEPTO the generation of the ME--partons
is based on the JADE algorithm. We generated the ME--partons with
the jet resolution parameter
$y_{\rm cut}={\rm PARL}(8)=0.008$. The parton jets and hadron
jets are determined with the higher value
$y_{\rm cut}=0.02$, in order to guarantee, that
the available phase space for the 2+1 jets of the jet algorithm~2 at
$y_{\rm cut}=0.02$ is (for the most part)
included in   the phase space for the 2+1 jets
of the jet algorithm~1 at $y_{\rm cut}=0.008$.
 (According to the PROJET Monte Carlo \cite{projet}
  contributions from 3+1 jets are small
   at $y_{\rm cut}=0.02$.)
We
order the parton jets and the hadron jets
 according to their distance to
the
proton $d_{\rm jet, prot}$.
The jet, which is closest to the proton direction, is considered as the remnant
jet, the next jet is called jet~1.

%
%

First we
concentrate on the hadron level
and compare the
transverse momentum ($p_T$) distributions of the jet 1
 for the jet algorithms 1 and 2
(grey distributions in figure 1a and 2a;
the hatched part
shows hadronization effects,
see later).
The $p_T$ distributions  differ strongly.
Both jet algorithms show up a maximum around $p_T = 10 \;GeV$,
but the jet algorithm 2 finds much more low $p_T$ jets,
i.e. the jet algorithm 2 is much more sensitive in the
low $p_T$ region, than the jet algorithm 1.
The different predictions of the low $p_T$--behavior
is caused by the mass of the jets,
although the
mean mass of the hadron jets is only of the order of 10~\%
of the reference
mass $M$.

The low $p_T$--behavior of the jet 1 is
dominated by hadronization effects.
This can be read off from the figures 1b and 2b, which
show the scatter plots of
the transverse momentum $p_T$ of the jet 1
(hadron level (HL) versus parton level (PL))
for the jet algorithm 1 and the jet algorithm 2, respectively.
{}From the figures 1b and 2b  follows moreover,
that the major part of the   hadronization effects is characterizable
by a transverse momentum $p_T({\rm HL})< 5\; GeV$.

The transverse momentum of the jets
is often used to control hadronization effects.
We suggest another method.
Consider the square of the distance
between the proton and jet1 scaled by the
square of the reference mass
\bea
      z^*_p = \frac{d^2_{\rm jet1,prot}}{M^2}.
\nonumber
\eea
The figures 1c and 2c show
the influence of
the cut
\bea
      z^*_p > 0.3
\label{cut}
\eea
on the scatter plots 1b and 2b.
As can be seen a lot of the offdiagonal events are removed
for both jet algorithms, i.e.
the cut (\ref{cut}) specifies a region in the phase space, where
parton jets can be considered as particle--like, at
least with respect to the transverse momentum of the jet 1.
A $p_T$--cut
 acts  ``horizontally'' in the figures 1b and 2b,
a  $z^*_p$--cut acts almost ``diagonally''.
Contrary to a $p_T$--cut, which depends on the reference frame,
a $z^*_p$--cut is Lorentz invariant and can always be applied
in the laboratory frame.
The hatched distributions in the figures 1a and 2a show the
transverse momentum of the jet 1 for those events, which do not fulfil
the $z_p^*$--cut (\ref{cut}). By a measurement of the
$p_T$--distribution of the jet 1, the increase below
$p_T=6\;{\rm GeV}$, predicted by the jet algorithm 2, might
be observable at HERA. Since the increase is caused by
hadronization effects, the jet algorithm 2 could be
useful to test hadronization models experimentally.

The scaled distance between the the jet1 and the remnant
$
 z_r^* = d^2_{\rm jet1, remnant}/ M^2
$
is less
suited practically to identify hadronization effects,
since most of the remnant escapes undetected in the beampipe.
While the momentum of the proton is known,
the momentum of the remnant can only be reconstructed
from the missing momentum in the detector.

Finally we show, that a $z_p^*$--cut is also useful
to control hadronization effects for 2+1 jets.
2+1 jets are important, since they can be used to measure
the strong coupling constant $\alpha_s$.
Let us consider the variable
\bea
     \eta =  \frac{Q^2_{\rm had}}{Q^2_{\rm had}
              + (p_{\rm jet1} +  p_{\rm jet2})^2 }  .
\nonumber
\eea
 The figures 3 and 4 show the $\eta$--scatter
 plots (hadron level versus parton level) for the
2+1 hadron jets of the
jet algorithms 1 and 2, respectively. In the figures 3a and 4a
no hadronization cuts are applied.
The correlations are very bad, indicating
strong hadronization effects for both jet algorithms.
After applying the $z^*_p$--cut (\ref{cut})
the correlations are considerably improved (figures 3b and 4b).
Since  strong $z^*_p$--cuts  lower the statistics,
one  has to look for a compromise between the
strength of the hadronization effects and the statistics of the data.
The ``best'' $z^*_p$--cut is the cut, at which
the uncertainty for the reconstruction of the parton level
from the hadron level is comparable to the uncertainty for
the reconstruction of the hadron level from the detector level.

Another important variable to characterize 2+1 jets
is
\bea
       z_p = \frac{p_{\rm prot}p_{\rm jet1}}{p_{\rm prot}q_{\rm had}}
\nonumber
\eea
Since in the limit of small masses
\bea
      z^*_p \rightarrow \frac{z_p}{1-x_{\rm Bj}}
\nonumber
\eea
one can consider $z^*_p$ as a generalization
of $z_p$ to massive jets.
The mean mass of hadron jets is of the order of 10~\%
of the reference
mass $M$ and therefore $z_p$ differs not very much from $z^*_p$
practically.
This means that
a measurement of $z_p$ cannot be performed below the $z^*_p$--cut.

In summary the hadronization effects of
the JADE algorithm and its simplest Lorentz invariant generalization
to massive jets  differ strongly around the direction of
the incoming proton.
By
adjusting the
$z^*_p$--cut properly the
differences disappear almost.
The hadronization effects from the proton remnant are controlable
without any special treatment of the proton remnant
during the clustering procedure.
The Lorentz invariant generalization of the JADE algorithm,
derived in section 2, is sensitive to hadronization
effects and might be useful in jet physics to test hadronization models
experimentally.

\section*{Acknowledgements}

We would like to thank D. Graudenz and H. Spiesberger for
many helpful and stimulating discussions.
We benefitted from conversations with
U. Bannier, T. Haas, J. Hartmann, G. Ingelman, G. Kramer,
K.~H. Rehren, S. Stiliaris, R.--D. Tscheuschner, P. Zerwas
and W. Zeuner.
The support from the ZEUS collaboration is acknowledged
with great appreciation.

\begin{figure}[htb]
\centerline{\psfig{figure=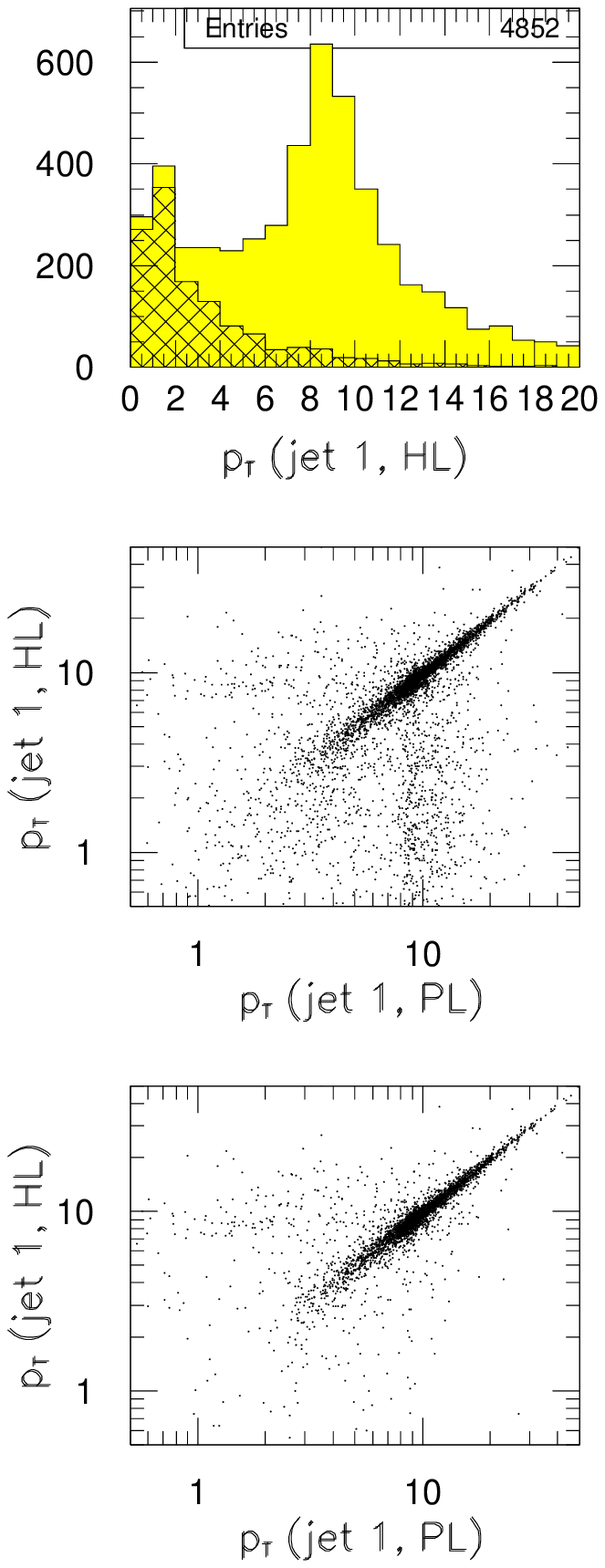,width=90mm,height=180mm}}
\caption{Jet algorithm 1: a)
           $p_T$ distribution (in [$GeV$]) of jet 1 (hadron level)
           in the hadronic center of mass frame.
           The hatched part shows
           hadronization effects with $z_p^*<0.3$.
  b) Scatter plot of the transverse momentum
$p_T$ (in [$GeV$]) of jet 1 in the hadronic center of mass frame:
 without a $z^*_p$--cut. c) As b) but with the cut $z^*_p>0.3$.}
\end{figure}

\begin{figure}[htb]
\centerline{\psfig{figure=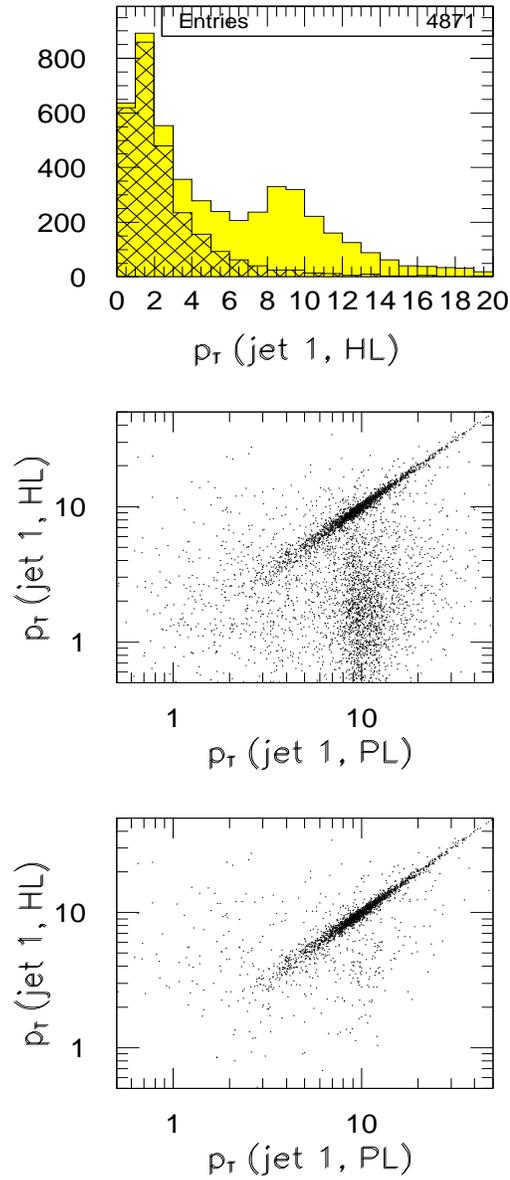,width=90mm,height=180mm}}
\caption{Jet algorithm 2:  a)
           $p_T$ distribution (in [$GeV$]) of jet 1 (hadron level)
           in the laboratory frame.
           The hatched part shows
           hadronization effects with $z_p^*<0.3$.
  b) Scatter plot of the transverse momentum
$p_T$ (in [$GeV$]) of jet 1 in the laboratory frame:
 without a $z^*_p$--cut. c) As b) but with the cut $z^*_p>0.3$.}
\end{figure}

\begin{figure}[htb]
\centerline{\psfig{figure=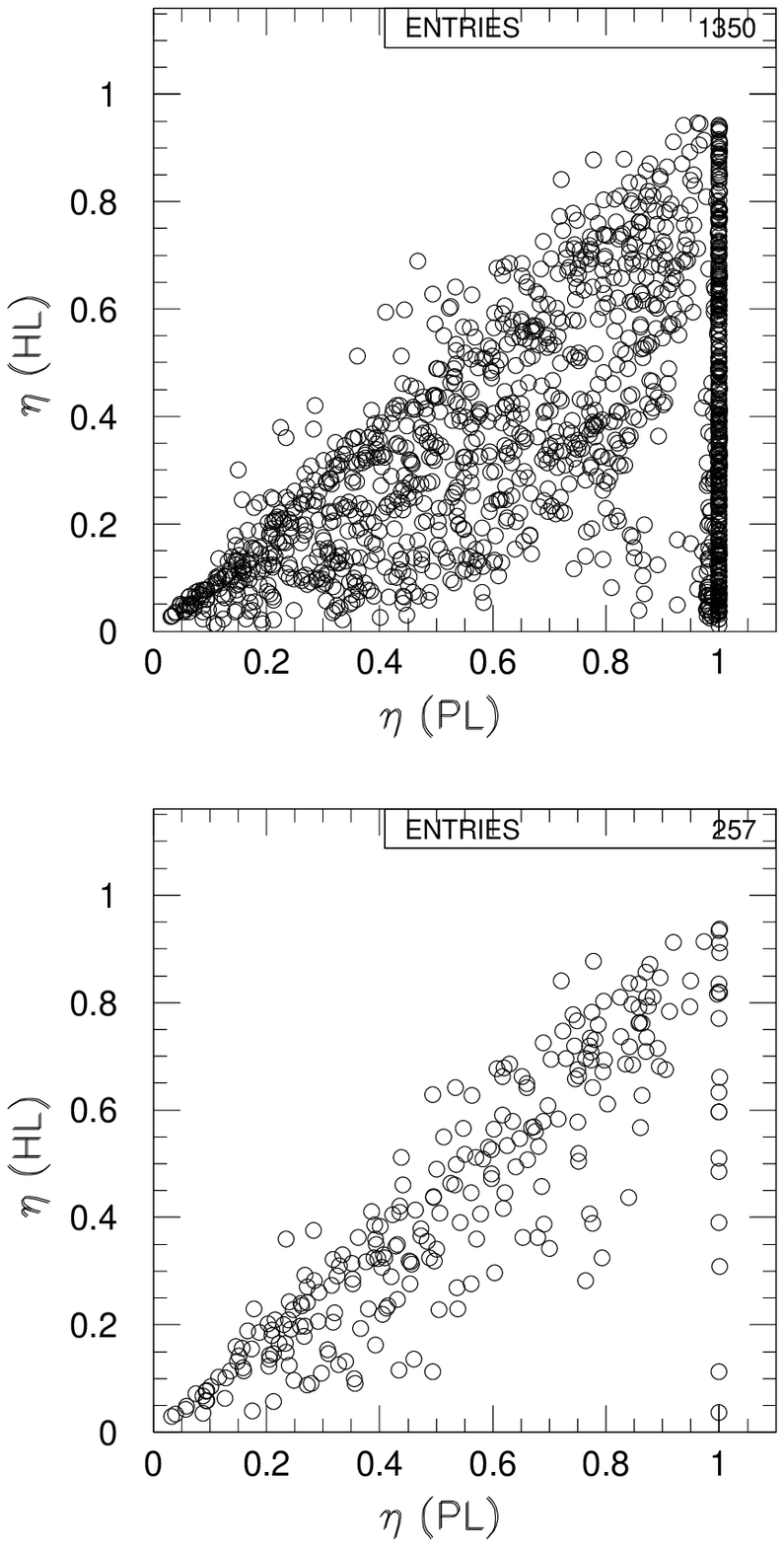,width=100mm,height=180mm}}
\caption{Jet algorithm 1:  scatter plot of $\eta$ for 2+1 hadron
                              jets: a) without a $z^*_p$--cut,
               b)  with the cut $z^*_p>0.3$.}
\end{figure}

\begin{figure}[htb]
\centerline{\psfig{figure=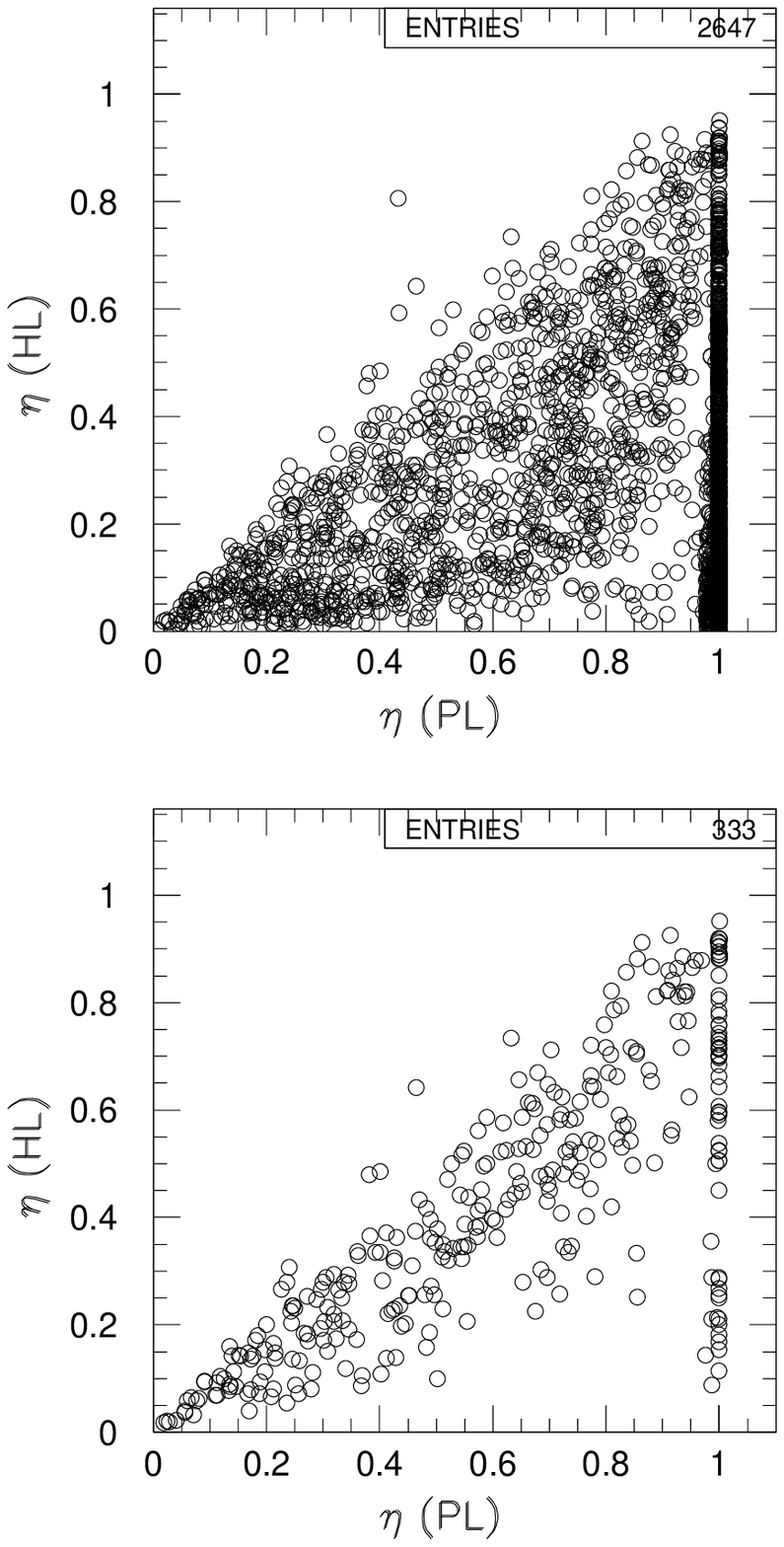,width=100mm,height=180mm}}
\caption{Jet algorithm 2:  scatter plot of $\eta$ for 2+1 hadron
                              jets: a) without   a $z^*_p$--cut,
               b)  with the cut $z^*_p>0.3$.}
\end{figure}

\end{document}